%% file: main.tex
\documentclass[conference]{IEEEtran}
\IEEEoverridecommandlockouts

\usepackage{graphicx}
\usepackage{tabularx}
\usepackage{enumitem}
\usepackage{siunitx} 
\usepackage{pifont}
\usepackage[font=small,skip=0pt]{caption}
\usepackage{ifthen}

\usepackage{array} %
\newcolumntype{P}[1]{>{\centering\arraybackslash}p{#1}} %
\usepackage{subcaption} 
\usepackage{comment}
\usepackage{xspace}

\usepackage{amsmath}

\usepackage{multirow}
\usepackage{makecell}

\usepackage{booktabs}
\usepackage{threeparttable} 
\usepackage[most]{tcolorbox}
\usepackage{listings}

\lstdefinestyle{mystyle}{
    basicstyle=\ttfamily\small, 
    breakatwhitespace=false,
    breaklines=true,
    captionpos=b, 
    frame=tb, 
    framesep=2pt,
    framerule=0pt,
    aboveskip=0pt,
    belowskip=0pt,
    xleftmargin=0pt, 
    xrightmargin=0pt, 
    numbers=none, 
}

\lstdefinestyle{javaStyle}{
  language=Java,
  basicstyle=\ttfamily\tiny,
  keywordstyle=\color{blue!70!black}\bfseries,
  stringstyle=\color{green!50!black},
  commentstyle=\color{gray}\itshape,
  numbers=none,
  frame=single,
  framerule=0.5pt,
  rulecolor=\color{gray},
  breaklines=true,
  tabsize=2,
  showstringspaces=false
}

\usepackage{xcolor}
\usepackage{etoolbox}

\usepackage{balance}

\usepackage{hyperref}
\hypersetup{
  colorlinks=true,
  linkcolor=transparent,
  anchorcolor=transparent,
  citecolor=black,
  urlcolor=black
}

\usepackage{cleveref}

\definecolor{mylightgray}{RGB}{224,224,224}

\newboolean{showcomments}
\setboolean{showcomments}{true}
\ifthenelse{\boolean{showcomments}}
 { \newcommand{\mynote}[2]{
      \fbox{\bfseries\sffamily\scriptsize#1}
        {\small$\blacktriangleright$\textsf{\emph{#2}}$\blacktriangleleft$}}}
        { \newcommand{\mynote}[2]{}}

\begin{document}

\title{Rethinking Cognitive Complexity for Unit Tests: Toward a Readability-Aware Metric Grounded in Developer Perception

\thanks{This research was funded by the Luxembourg National Research Fund (FNR), grant reference AFR PhD bilateral, project reference 17185670. Yinghua Li is the corresponding author.}
}

\author{\IEEEauthorblockN{1\textsuperscript{st} Wendkûuni C. Ouédraogo}
\IEEEauthorblockA{ 
\textit{University of Luxembourg}\\
wendkuuni.ouedraogo@uni.lu}
\and
\IEEEauthorblockN{2\textsuperscript{nd} Yinghua Li}
\IEEEauthorblockA{
\textit{University of Luxembourg}\\
yinghua.li@uni.lu}
\and
\IEEEauthorblockN{3\textsuperscript{rd} Xueqi Dang}
\IEEEauthorblockA{
\textit{University of Luxembourg}\\
xueqi.dang@uni.lu}
\and
\IEEEauthorblockN{4\textsuperscript{th} Xin Zhou}
\IEEEauthorblockA{
\textit{Singapore Management University}\\
xinzhou.2020@phdcs.smu.edu.sg}
\and
\IEEEauthorblockN{5\textsuperscript{th} Anil Koyuncu}
\IEEEauthorblockA{
\textit{Bilkent University}\\
anil.koyuncu@cs.bilkent.edu.tr}
\and
\IEEEauthorblockN{6\textsuperscript{th} Jacques Klein}
\IEEEauthorblockA{
\textit{University of Luxembourg}\\
jacques.klein@uni.lu}
\and
\IEEEauthorblockN{7\textsuperscript{th} David Lo}
\IEEEauthorblockA{
\textit{Singapore Management University}\\
davidlo@smu.edu.sg}
\and
\IEEEauthorblockN{8\textsuperscript{th} Tegawendé F. Bissyandé}
\IEEEauthorblockA{
\textit{University of Luxembourg}\\
tegawende.bissyande@uni.lu}
}

\maketitle

\input{./Sections/abstract.tex}


\input{./Sections/intro.tex}
\input{./Sections/motivations.tex}
\input{./Sections/empiricalObservations.tex}
\input{./Sections/limitationOfCurrentApproach.tex}
\input{./Sections/proposal.tex}
\input{./Sections/discussion.tex}
\input{./Sections/relatedWork.tex}

\input{./Sections/conclusion.tex}



\balance
\bibliographystyle{IEEEtran}
\bibliography{references}

\end{document}

%% file: Sections/abstract.tex
\begin{abstract}
Automatically generated unit tests—from search-based tools like EvoSuite or LLMs—vary significantly in structure and readability. Yet most evaluations rely on metrics like Cyclomatic Complexity and Cognitive Complexity, designed for functional code rather than test code.
Recent studies have shown that SonarSource’s Cognitive Complexity metric assigns near-zero scores to LLM-generated tests, yet its behavior on EvoSuite-generated tests and its applicability to test-specific code structures remain unexplored.
We introduce \textbf{CCTR}, a \textit{Test-Aware Cognitive Complexity} metric tailored for unit tests. CCTR integrates structural and semantic features like assertion density, annotation roles, and test composition patterns—dimensions ignored by traditional complexity models but critical for understanding test code.
We evaluate 15,750 test suites generated by EvoSuite, GPT-4o, and Mistral Large-1024 across 350 classes from Defects4J and SF110. Results show CCTR effectively discriminates between structured and fragmented test suites, producing interpretable scores that better reflect developer-perceived effort. By bridging structural analysis and test readability, CCTR provides a foundation for more reliable evaluation and improvement of generated tests. We publicly release all data, prompts, and evaluation scripts to support replication.
\end{abstract}

%% file: Sections/intro.tex
\section{Introduction}
\label{introduction}

As test automation evolves, traditional tools such as EvoSuite~\cite{fraser2011evosuite} are increasingly complemented by Large Language Model (LLM)-based approaches~\cite{siddiq2024using, tang2024chatgpt, chen2024chatunitest, wang2024software, ouedraogo2024llms, yang2024evaluation, ouedraogo2024large}, enabling scalable unit test generation across a wide range of software systems.
However, recent studies report that only 40–70\% of LLM-generated unit tests compile successfully without manual interventions~\cite{yang2024evaluation, tang2024chatgpt, siddiq2024using, ouedraogo2024large}. Even when syntactically correct, the structural quality and readability of these tests remain inconsistent and difficult to evaluate.

This raises the question of how to meaningfully assess the complexity and maintainability of generated test code—an area where existing metrics offer limited guidance. Traditional static complexity metrics like Cyclomatic Complexity~\cite{mccabe1976complexity} and SonarSource's Cognitive Complexity~\cite{sonarsource_complexity} were originally designed to evaluate functional logic, not test code. 
While recent studies~\cite{biagiola2024improving,deljouyi2024leveraging,tang2024chatgpt,ouedraogo2024llms,ouedraogo2024large} consistently report very low cognitive complexity scores for LLM-generated tests, none have evaluated this metric on EvoSuite-generated suites or examined its ability to reflect the structural and semantic complexity of test code—raising concerns about their suitability for assessing test quality.

In parallel, research on test readability~\cite{winkler2024investigating, daka2015modeling} highlights the importance of assertion clarity, descriptive naming, and structural conventions. The Arrange–Act–Assert (AAA) pattern, in particular, has been widely adopted for improving test readability by clearly separating setup, execution, and verification phases. The machine-learned readability model proposed by Scalabrino et al.~\cite{scalabrino2018comprehensive}, and later validated by Sergeyuk et al.~\cite{sergeyuk2024reassessing}, approximates human judgments of code readability based on textual features. However, it focuses primarily on surface-level attributes—such as identifier clarity and formatting—and does not account for deeper structural aspects like fragmentation, assertion density, or control-flow complexity. This leaves a critical gap in the quantitative assessment of test structure and maintainability.


We introduce \textbf{CCTR}, a \textit{Test-Aware Cognitive Complexity} metric tailored for unit tests. CCTR integrates structural elements—assertion usage, annotation roles, and test composition patterns—into a lightweight, interpretable score that reflects both the syntactic structure and the congitive effort required to understand test suites. 
To assess its effectiveness, we evaluate 15,750 test suites generated by EvoSuite, GPT-4o, and Mistral Large across Defects4J~\cite{just2014defects4j} and SF110~\cite{panichella2017automated}. Results show that CCTR effectively discriminates between fragmented, auto-generated tests and well-structured code, aligning more closely with dimensions valued by developers—such as clarity of assertions, test structure, and semantic intent—compared to SonarSource’s Cognitive Complexity~\cite{sonarsource_complexity}, while providing a meaningful proxy for test maintainability.



\vspace{0.5em}
\noindent\textbf{Contributions.} This paper provides:
\begin{itemize}
    \item A critical analysis of the limitations of Cognitive Complexity when applied to test code;
    \item A novel metric, CCTR, designed to quantify structural and semantic complexity in unit tests;
    \item A large-scale empirical evaluation of CCTR across 15,750 test suites;
    \item Evidence demonstrating that CCTR more accurately reflects test readability aspects compared to existing metrics;
    \item A publicly available dataset of 15,750 generated test suites across 350 Java classes, including CCTR scores, prompt templates, and evaluation scripts: \url{https://github.com/Cwendkuuni/CCTR}
\end{itemize}

\vspace{0.5em}
\noindent\textbf{Paper organization.}
The remainder of this paper is structured as follows: Section~\ref{motivating_example} motivates the need for a test-specific metric through concrete examples; Section~\ref{empirical_observations_at_scale} presents our experimental setup and dataset; Section~\ref{limitations_of_cognitive_complexity} discusses the shortcomings of existing complexity metrics; Section~\ref{Toward_a_test_specific_cognitive_complexity} introduces and evaluates CCTR; Section~\ref{discussion} discusses findings and limitations; Section~\ref{related_work} reviews related work; and Section~\ref{conclusion} concludes.

%% file: Sections/motivations.tex
\section{Motivating Example}
\label{motivating_example}

%
%
To motivate the need for a test-specific complexity metric, we present three examples: two real test suites for the \texttt{CommandLine} class from the \texttt{Cli-1} bug (Defects4J), and one synthetic control-flow case. Full versions are available in our replication package. We use PMD’s~\cite{pmd} implementation of Cognitive Complexity, based on SonarSource’s specification~\cite{sonarsource_complexity}, which scores control-flow constructs and nesting to estimate mental effort.
%

\vspace{-3mm}
\subsection{Example 1: Deeply Nested Trivial Code}

\begin{lstlisting}[style=javaStyle, caption={Manually written nested code}, label=lst:nested_example]
public class ExampleTest {
    public void testExample() {
        if (condition) {
            for (int i = 0; i < 10; i++) {
                while (anotherCondition) {
                    // nested logic
                }
            }
        }
    }
}
\end{lstlisting}

Listing~\ref{lst:nested_example} exhibits structural nesting but minimal logic. PMD assigns it a Cognitive Complexity of 12.

\subsection{Example 2: LLM-Generated Test Suite}

\begin{lstlisting}[style=javaStyle, caption={Excerpt from GPT-4o-generated test}, label=lst:llm_test]
@Test
public void testGetOptionValueWithDefaultValue() {
    assertEquals("default", commandLine.getOptionValue("test", "default"));
}
\end{lstlisting}

Listing~\ref{lst:llm_test} shows part of a 12-method suite generated by GPT-4o. Despite meaningful naming, consistent structure, and annotations, comprehension still requires effort due to multiple methods, assertion logic, and parameter variation. Yet PMD assigns it a Cognitive Complexity of 0, missing this moderate structural and semantic load.

\subsection{Example 3: EvoSuite-Generated Test Suite}

\begin{lstlisting}[style=javaStyle, caption={Excerpt from EvoSuite-generated test}, label=lst:evosuite_test]
@Test(timeout = 4000)
public void test02() throws Throwable {
    CommandLine commandLine0 = new CommandLine();
    Option option0 = new Option("", "\"Jd", true, "D1,L");
    option0.addValue("org.apache.commons.cli.CommandLine");
    commandLine0.addOption(option0);
    String string0 = commandLine0.getOptionValue("--", null);
    assertNotNull(string0);
    assertEquals("org.apache.commons.cli.CommandLine", string0);
}
\end{lstlisting}

Listing~\ref{lst:evosuite_test} is part of a 17-method suite generated by EvoSuite. It features cryptic naming and lacks structural clarity or semantic grouping. Yet, its Cognitive Complexity score—according to PMD—is 0. \\

As these examples illustrate, the control-flow-oriented approach fails to capture key structural and semantic aspects of unit tests. Despite clear differences in readability, structure, and developer intent, all three cases receive nearly identical scores—highlighting a misalignment between the SonarSource assumptions and the actual comprehension effort involved in test code.

\subsection{Summary}

Table~\ref{tab:motivating_scores} summarizes the PMD Cognitive Complexity scores across the three examples. To better illustrate the disconnect between PMD’s scoring and intuitive comprehension effort, we include a perceived complexity label based on structural and semantic clarity. While the trivial nested example receives a non-zero score, more cognitively demanding test suites are rated as zero—highlighting a critical misalignment with developer intuition.

\begin{table}[ht]
\centering
\caption{Cognitive Complexity (PMD) vs. Perceived Complexity}
\label{tab:motivating_scores}
\scalebox{0.7}{
\begin{tabular}{lccc}
\textbf{Example} & \textbf{Origin} & \textbf{PMD Score} & \textbf{Perceived Complexity} \\
\hline
Nested Trivial Code     & Manual    & 12 & Low \\
LLM-Generated Test Suite & GPT-4o    & 0  & Medium \\
EvoSuite Test Suite      & EvoSuite  & 0  & High \\
\end{tabular}
}
\vspace{-3mm}
\end{table}

These results highlight a critical gap: while PMD captures structural nesting, it fails to reflect the human-perceived complexity of real-world test suites—especially those with rich assertion logic, mocking, or ambiguous structure. This motivates the introduction of a test-specific cognitive complexity metric in Section~\ref{Toward_a_test_specific_cognitive_complexity}.

%% file: Sections/empiricalObservations.tex
\section{Empirical Observations at Scale}
\label{empirical_observations_at_scale}

\subsection{Dataset and Test Generation Setup}
\label{dataset_and_generated_tests}

To evaluate CCTR, we conducted a large-scale study using two established Java code corpora: Defects4J~\cite{just2014defects4j} and SF110~\cite{panichella2017automated}. These benchmarks provide diverse real-world classes with varying complexity, serving as a robust foundation for evaluating both traditional and LLM-generated test suites.

\noindent\textbf{Dataset composition.}
We selected 15 projects from Defects4J (147 classes) and 69 projects from SF110 (203 classes), totaling 350 classes across 84 projects. Table~\ref{tab:dataset_stats} summarizes the overall characteristics.

\begin{table}[ht]
\centering
\caption{Dataset Composition and Code Statistics}
\label{tab:dataset_stats}
\scalebox{0.7}{
\begin{tabular}{lcccccc}
\textbf{Dataset} & \textbf{\#Projects} & \textbf{\#Classes} & \textbf{LOC avg} & \textbf{Token avg} & \textbf{Methods/Class avg} \\
\hline
Defects4J        & 15                  & 147                & 127.76           & 1779.67             & 11.88                      \\
SF110            & 69                  & 203                & 146.82           & 1518.86             & 14.21                      \\
\end{tabular}
}
\vspace{-3mm}
\end{table}

\noindent\textbf{Test generation.}
We generated unit tests for each class using three approaches: EvoSuite~\cite{fraser2011evosuite} with a 3-minute time budget and the DynaMOSA search algorithm~\cite{panichella2017automated}, GPT-4o~\cite{hurst2024gpt}, and Mistral Large-2407\footnote{\url{https://mistral.ai/news/mistral-large-2407}} with zero-shot prompting. To mitigate randomness in LLM outputs, we set the temperature to 0.1 and kept top-p at its default. Each tool/model was run 15 times per class, resulting in 2,205 test suites per model on Defects4J and 3,045 per model on SF110. In total, we produced 10,500 test suites with LLMs and 5,250 with EvoSuite, yielding 15,750 test suites overall.

\noindent\textbf{Test characteristics.}
Table~\ref{tab:test_stats} presents average metrics across all generated tests. While EvoSuite produced longer test classes with more methods and higher token counts, LLM-generated tests were more concise yet semantically expressive. Compilation rates were 100\% for EvoSuite and lower for LLMs due to occasional syntax or API misuses.

\begin{table}[ht]
\centering
\caption{Generated Test Suite Statistics per Dataset and Model}
\label{tab:test_stats}
\scalebox{0.7}{
\begin{tabular}{llccccc}
\textbf{Dataset} & \textbf{Model} & \textbf{\#Tests} & \textbf{LOC avg} & \textbf{Token avg} & \textbf{Methods avg} & \textbf{Compilability} \\
\hline
\multirow{3}{*}{Defects4J} 
  & EvoSuite             & 2205 & 169.40 & 1958.19 & 17.38 & 100\% \\
  & GPT-4o               & 2205 & 95.04  & 822.98  & 14.45 & 67.58\% \\
  & Mistral-L        & 2205 & 112.06 & 939.33  & 16.20 & 42.42\% \\
\hline
\multirow{3}{*}{SF110} 
  & EvoSuite             & 3045 & 294.14 & 3462.57 & 27.82 & 100\% \\
  & GPT-4o               & 3045 & 96.40  & 894.36  & 13.71 & 49.80\% \\
  & Mistral-L        & 3045 & 112.06 & 939.33  & 16.20 & 42.42\% \\
\end{tabular}
}
\vspace{-3mm}
\end{table}


\subsection{Code Complexity and Readability Analysis}

We present a large-scale comparative analysis of unit test suites generated by LLMs and EvoSuite across three complementary perspectives: Cyclomatic Complexity, Cognitive Complexity, and automated readability scores from Scalabrino et al.~\cite{scalabrino2018comprehensive}.

\vspace{0.5em}
\noindent\textbf{Cyclomatic complexity.}
Cyclomatic Complexity~\cite{mccabe1976complexity} estimates linearly independent paths through code and approximates control-flow density. As shown in Table~\ref{tab:cyclomatic}, test suites generated by LLMs and EvoSuite exhibit substantial control-flow complexity, particularly in SF110, confirming that generated test code contains non-trivial execution paths and is far from structurally simplistic.

\begin{table}[ht]
\centering
\caption{Cyclomatic Complexity and Readability Scores across Models for Each Dataset.}
\label{tab:cyclomatic_readability_grouped}

\subfloat[\textbf{Cyclomatic Complexity}\label{tab:cyclomatic}]{
\scalebox{0.7}{
\begin{tabular}{llrrrrrr}
 \toprule
\textbf{Dataset} & \textbf{Model/Tool} & \textbf{Min} & \textbf{Q1} & \textbf{Median} & \textbf{Q3} & \textbf{Max} & \textbf{Mean} \\
\midrule
\multirow{3}{*}{Defects4J}
  & GPT-4o         & 2 & 10 & 13 & 19 & 35  & 14.95 \\
  & Mistral-L      & 2 & 10 & 12 & 18 & 37  & 14.89 \\
  & EvoSuite       & 1 &  6 & 17 & 29 & 131 & 19.78 \\
\midrule
\multirow{3}{*}{SF110}
  & GPT-4o         & 3 &  8 & 14 & 19 & 92  & 15.69 \\
  & Mistral-L      & 3 &  9 & 13 & 19 & 92  & 16.91 \\
  & EvoSuite       & 1 &  8 & 17 & 34 & 230 & 25.55 \\
\bottomrule
\end{tabular}
}
}

\vspace{1em}

\subfloat[\textbf{Readability Scores (Scalabrino et al., 0--100)}\label{tab:readability}]{
\scalebox{0.7}{
\begin{tabular}{llrrrrrr}
 \toprule
\textbf{Dataset} & \textbf{Model/Tool} & \textbf{Min} & \textbf{Q1} & \textbf{Median} & \textbf{Q3} & \textbf{Max} & \textbf{Mean} \\
\midrule
\multirow{3}{*}{Defects4J}
  & GPT-4o         & 0.0  & 63.0 & 69.3 & 74.2 & 87.5 & 67.0 \\
  & Mistral-L      & 19.2 & 62.3 & 69.0 & 74.0 & 85.2 & 67.0 \\
  & EvoSuite       & 8.0  & 41.8 & 58.5 & 72.8 & 95.7 & 56.3 \\
\midrule
\multirow{3}{*}{SF110}
  & GPT-4o         & 23.8 & 58.7 & 66.3 & 70.8 & 90.3 & 64.4 \\
  & Mistral-L      & 24.6 & 60.2 & 67.3 & 72.4 & 84.4 & 64.9 \\
  & EvoSuite       & 4.1  &   -- & 50.9 &   -- & 95.7 & 51.1 \\
\bottomrule
\end{tabular}
}
}
\vspace{-3mm}
\end{table}

\vspace{0.5em}
\noindent\textbf{Readability.}
To approximate perceived readability at scale, we use the model by Scalabrino et al.~\cite{scalabrino2018comprehensive}, which combines 104 structural, textual, and visual features trained on annotated Java code, including JUnit tests. As shown in Table~\ref{tab:readability}, it consistently ranks LLM-generated tests as more readable than those from EvoSuite. Sergeyuk et al.~\cite{sergeyuk2024reassessing} found this model to best align with human ratings among established alternatives~\cite{buse2008metric,buse2009learning,posnett2011simpler,dorn2012general,mi2022towards}.

\vspace{0.5em}
\noindent\textbf{Cognitive complexity.}
We use PMD's implementation of Cognitive Complexity~\cite{sonarsource_complexity}, which penalizes nested control structures, logical operators, and flow-disrupting elements. As shown in Table~\ref{tab:sonar_complexity}, the metric yields extremely low values—often zero—for LLM-generated tests and modest scores for EvoSuite despite highly fragmented test classes. This suggests that Cognitive Complexity overlooks many characteristics that increase cognitive effort in understanding unit tests, including assertion logic, mocking, and test-specific design idioms.

\begin{table}[t]
\centering
\caption{Sonar Cognitive Complexity (PMD).}
\label{tab:sonar_complexity}
\scalebox{0.7}{
\begin{tabular}{llrrrrrr}
\textbf{Dataset} & \textbf{Model/Tool} & \textbf{Min} & \textbf{Q1} & \textbf{Median} & \textbf{Q3} & \textbf{Max} & \textbf{Mean} \\
\hline
\multirow{3}{*}{Defects4J}
  & GPT-4o         & 0 & 0 & 0 & 0 & 8   & 0.285 \\
  & Mistral-L  & 0 & 0 & 0 & 0 & 14  & 0.761 \\
  & EvoSuite       & 0 & 1 & 2 & 6 & 30  & 3.931 \\
\hline
\multirow{3}{*}{SF110}
  & GPT-4o         & 0 & 0 & 0 & 0 & 46  & 0.585 \\
  & Mistral-L  & 0 & 0 & 0 & 0 & 46  & 1.203 \\
  & EvoSuite       & 0 & 1 & 2 & 6 & 102 & 4.745 \\
 \bottomrule
\end{tabular}
}
\vspace{-3mm}
\end{table}

\vspace{0.5em}
\noindent\textbf{Takeaway.}
Each perspective captures a specific facet: control-flow density (Cyclomatic), perceived readability (Scalabrino), and nesting-driven logic effort (Cognitive Complexity). Cyclomatic and Scalabrino operate effectively within their scopes, but Sonar's Cognitive Complexity—despite growing use in quality assessments—was designed for functional code, not test code. Our results show it fails to reflect structural and semantic effort in understanding unit tests, overlooking test-specific patterns like assertion logic, mocking constructs, and annotation-based roles. This gap motivates the need for a dedicated cognitive complexity metric that reflects the unique characteristics of test code. 

%% file: Sections/limitationOfCurrentApproach.tex
\section{Limits of Cognitive Complexity for Tests}
\label{limitations_of_cognitive_complexity}

Although SonarSource's Cognitive Complexity metric~\cite{sonarsource_complexity} is widely adopted to assess source code understandability, it was originally designed to measure the mental effort required to comprehend functional logic—not unit tests. In this section, we identify three key limitations that emerge when applying this metric to test code.

\vspace{0.5em}
\subsection{Not Designed for Unit Test Constructs}

The Cognitive Complexity metric penalizes nested control structures (\texttt{if}, \texttt{for}, \texttt{while}), logical operators, and flow interruptions (\texttt{break}, \texttt{return})—constructs common in business logic but less prevalent in well-written unit tests. Conversely, constructs heavily influencing test code readability and structure—assertions, mocking invocations, and annotations like \texttt{@Test} or \texttt{@ParameterizedTest}—are completely ignored. Test methods that heavily rely on test-specific constructs—such as assertions, mocking invocations, and annotations—may still receive zero complexity scores, undermining the metric’s applicability for evaluating generated tests or comparing test suites across tools.

\vspace{0.5em}
\subsection{Empirical Observations from Generated Tests}

We report empirical observations from our analysis of 15,750 test suites generated by EvoSuite, GPT-4o, and Mistral Large, as previously detailed in Sections~\ref{motivating_example} and~\ref{empirical_observations_at_scale}. Despite significant differences in structural quality and readability, PMD's Cognitive Complexity assigns zero scores to over 99\% of LLM-generated test methods and only modest scores to EvoSuite outputs. As illustrated in Section~\ref{motivating_example}, even a deeply nested dummy example scored 12, while real test suites containing semantically rich and non-trivial assertions often received a score of zero. This highlights a key limitation: the metric fails to capture the structural and cognitive effort inherent in test code, particularly for tests generated by modern LLMs.

\subsection{Lack of Alignment with Test Readability Dimensions}

Recent studies emphasize that test code readability extends beyond control flow. Winkler et al.~\cite{winkler2024investigating} show developers prioritize clear structure (AAA pattern), meaningful naming, assertion logic, and setup clarity. Similarly, Daka et al.~\cite{daka2015modeling} and Biagiola et al.~\cite{biagiola2024improving} highlight that human-perceived test quality correlates more with naming, structure, and assertion intent than nesting or branches. Yet none of these elements are captured by Cognitive Complexity. The increasing use of this metric for evaluating LLM-generated tests~\cite{biagiola2024improving,deljouyi2024leveraging} lacks foundational justification. A test-specific complexity metric should reflect cognitive effort from test-specific constructs—not control-flow alone.

\vspace{0.5em}
\noindent These limitations motivate the development of a new metric that accounts for the actual structural and semantic features of test suites. 


%% file: Sections/proposal.tex
\section{Toward a Test-Specific Cognitive Complexity}
\label{Toward_a_test_specific_cognitive_complexity}


We propose \textbf{CCTR} (Cognitive Complexity for Test Readability), a test-specific metric designed to quantify structural and semantic effort in test comprehension. CCTR addresses limitations in existing metrics by capturing constructs that developers actively process when reading tests—assertions, mocking patterns, and annotations—while maintaining control-flow sensitivity.

\subsection{Empirical Motivation}


%
CCTR is grounded in the developer-centric empirical model proposed by Winkler et al.~\cite{winkler2024investigating}, which identifies three main dimensions of test comprehension: \textbf{test structure} (e.g., AAA clarity, indentation), \textbf{naming and expression semantics} (e.g., assertion intent, variable clarity), and \textbf{test logic and purpose} (e.g., relevance of setup, test goal). Each component of CCTR is designed to reflect one or more of these dimensions. In particular, our inclusion of annotations is inspired by Guerra et al.~\cite{guerra2024annotations}, who found that annotations like \texttt{@Test} and \texttt{@NotNull} improve code readability and convey test roles—but also warned against overuse. CCTR assigns weights accordingly, balancing informativeness with complexity impact.

\subsection{CCTR Definition}

We define CCTR as:
\begin{equation}
\mathrm{CCTR} = \alpha \cdot N + \beta \cdot A + \gamma \cdot M + \delta \cdot T
\end{equation}

\noindent Where $N$ denotes the control-flow nesting complexity, following the original SonarSource definition. The term $A$ corresponds to the number of assertion or \texttt{fail()} statements present in the test method. The component $M$ captures the number of mocking-related constructs, such as calls to \texttt{mock()}, \texttt{verify()}, or \texttt{when()}. Finally, $T$ represents annotation-based signaling. It is computed by assigning a value of +1 for each occurrence of common test-related annotations such as \texttt{@Test}, \texttt{@BeforeEach}, and \texttt{@AfterEach}, and a value of +2 for the presence of more specialized annotations such as \texttt{@ParameterizedTest}. \\

We adopt initial weights:
\[
\alpha = \beta = \gamma = \delta = 1.0
\]

This uniform weighting reflects our design goal of simplicity and interpretability, while aligning with empirical studies suggesting these factors contribute comparably to test comprehension. Our treatment of annotations is further informed by Guerra et al.~\cite{guerra2024annotations}, who show that annotations like \texttt{@Test} and \texttt{@NotNull} enhance structural clarity and intent signaling—especially when defining test roles or validation rules. However, they also caution against overuse or ambiguity. To balance informativeness and score inflation, CCTR assigns moderate, fixed weights to annotation types. Tuning these weights remains future work guided by large-scale developer studies.

\subsection{Evaluation and Observations}

To evaluate CCTR, we computed scores across two benchmarks (Defects4J, SF110) and three test generation strategies: GPT-4o, Mistral Large, and EvoSuite.

\begin{table}[t]
\centering
\caption{CCTR (Test-Aware Cognitive Complexity)}
\label{tab:cctr}
\scalebox{0.7}{
\begin{tabular}{llrrrrrr}
\textbf{Dataset} & \textbf{Model/Tool} & \textbf{Min} & \textbf{Q1} & \textbf{Median} & \textbf{Q3} & \textbf{Max} & \textbf{Mean} \\
\hline
\multirow{3}{*}{Defects4J}
  & GPT-4o         & 3 & 13 & 23 & 34 & 127  & 26.44 \\
  & Mistral-L  & 4 & 13 & 23 & 40 & 132  & 27.91 \\
  & EvoSuite       & 0 & 10 & 31 & 53 & 494  & 39.47 \\
\midrule
\multirow{3}{*}{SF110}
  & GPT-4o         & 2 & 16 & 24 & 34 & 206  & 28.41 \\
  & Mistral-L  & 2 & 16 & 23 & 34 & 206  & 30.61 \\
  & EvoSuite       & 0 & 15 & 38 & 75 & 1557 & 58.61 \\
 \bottomrule
\end{tabular}
}
\vspace{-3mm}
\end{table}

\vspace{0.5em}

%
\noindent\textbf{Results.} Several insights emerge from Table~\ref{tab:cctr}. \textbf{LLM-generated test suites} (GPT-4o, Mistral) show moderate complexity (mean 26–30), consistent with their structured yet concise test style. In contrast, \textbf{EvoSuite tests} score significantly higher due to their fragmented structure, high method count, and dense assertions. CCTR also distinguishes between test suites that SonarSource’s Cognitive Complexity fails to separate—capturing both generator-level and dataset-level differences. Finally, despite variability in LOC and token count, CCTR scales robustly across tools and datasets. These results support that CCTR better reflects test comprehension effort, aligning with developer-centric principles and structural variation.

\subsection{Illustration}

To illustrate how CCTR captures structural and semantic variation, we revisit the three examples from Section~\ref{motivating_example}. Table~\ref{tab:illustration_examples} shows both PMD and CCTR scores, highlighting how traditional metrics fail to reflect comprehension differences. 

\begin{table}[ht]
\centering
\caption{Comparison of PMD and CCTR scores across example test classes}
\label{tab:illustration_examples}
\scalebox{0.7}{
\begin{tabular}{lccc}
\textbf{Example} & \textbf{PMD Score} & \textbf{CCTR Score} & \textbf{Perceived Complexity} \\
\hline
Nested Trivial Code     & 12  & 12  & Low \\
LLM-Generated Test Suite & 0   & 12  & Medium \\
EvoSuite Test Suite      & 0   & 35  & High \\
 \bottomrule
\end{tabular}
}
\vspace{-3mm}
\end{table}




%
Each test class represents a distinct style and level of structural clarity. \textbf{Nested Trivial Code} (Listing~\ref{lst:nested_example}), while semantically simple, involves deep control-flow nesting and yields a moderate CCTR score of \textbf{12}, reflecting structural complexity. \textbf{LLM-Generated Test Suite} (Listing~\ref{lst:llm_test}) achieves the same score, but for different reasons: concise methods, consistent assertions, clear naming, and informative annotations. In contrast, the \textbf{EvoSuite-Generated Test Suite} (Listing~\ref{lst:evosuite_test}) contains 17 fragmented methods with synthetic names and redundant assertions, resulting in a much higher CCTR score of \textbf{35}, indicative of increased comprehension effort.

These examples show that CCTR produces meaningful, differentiated scores at the class level, reflecting structural variation across test styles. Combined with Table~\ref{tab:cctr}, this illustrates its applicability to both individual cases and large-scale benchmarks.

%% file: Sections/discussion.tex
\section{Discussion and Future Work}
\label{discussion}


\noindent\textbf{Implications for practice.}
CCTR offers a test-specific lens for evaluating unit tests where readability matters, aiding refactoring, design, and automated review. By integrating assertions, mocking, and annotations, it provides a cognitively grounded proxy for maintainability. CCTR also scales to auto-generated tests, enabling large-batch evaluation of synthesis quality.

\vspace{0.5em}
\noindent\textbf{Revisiting prior metrics.}
Our results reveal that existing structural metrics like Cyclomatic Complexity and Cognitive Complexity fail to reflect semantic and structural dimensions central to unit test design. This explains why prior studies~\cite{biagiola2024improving,deljouyi2024leveraging} relied on workarounds like CodeBLEU~\cite{ren2020codebleu} or cosine similarity for refactoring validation despite lacking theoretical grounding. CCTR offers a principled alternative.

\vspace{0.5em}
\noindent\textbf{Limitations.}
CCTR, like other static metrics, does not capture all test quality aspects—it misses behavioral correctness, flakiness, and fault detection. Its fixed weights, though empirically motivated, may not generalize across all test styles. Still, CCTR complements dynamic analyses and other metrics to support broader assessment. Cross-domain and cross-language validation remains future work.

\vspace{0.5em}
\noindent\textbf{Future work.}
Several extensions are planned. First, we aim to conduct a human study to validate the correlation between CCTR scores and developer-perceived test comprehensibility. Second, we plan to explore CCTR’s utility in test generation workflows, particularly as a scoring signal for selecting or prioritizing LLM-generated test candidates. Finally, we envision extending CCTR into a multi-dimensional model that incorporates code coverage and fault-detection ability to offer a more comprehensive view of test effectiveness.

%% file: Sections/relatedWork.tex
\section{Related Work}
\label{related_work}


\noindent\textbf{Code complexity metrics.}
McCabe’s Cyclomatic Complexity~\cite{mccabe1976complexity} models control-flow density, while Cognitive Complexity~\cite{sonarsource_complexity} penalizes nesting and flow disruptions to approximate mental effort. Yet, empirical studies~\cite{munoz2020empirical,sharma2020we} show limited alignment with human comprehension. Our findings confirm that PMD’s implementation fails to distinguish test classes with different structural and semantic clarity. In contrast, CCTR captures test-specific constructs—assertions, mocking, annotations—offering a better proxy for perceived test complexity.

\vspace{0.5em}
\noindent\textbf{Readability models.}
Several models estimate code readability using lexical, syntactic, or visual features, including work by Buse and Weimer~\cite{buse2008metric,buse2009learning}, Dorn et al.~\cite{dorn2012general}, and Posnett et al.~\cite{posnett2011simpler}. Scalabrino et al.~\cite{scalabrino2018comprehensive} proposed a 104-feature model, later validated by Sergeyuk et al.~\cite{sergeyuk2024reassessing}. While effective for general code, these models emphasize surface-level cues and overlook structural or cognitive traits crucial to test code. CCTR addresses this gap with a metric tailored to test-specific comprehension.

\vspace{0.5em}
\noindent\textbf{Evaluating generated tests.}
With LLM-based code generation, new methods assess generated test quality. Biagiola et al.\cite{biagiola2024improving} and Deljouyi et al.\cite{deljouyi2024leveraging} used CodeBLEU and semantic similarity to detect hallucinations or assess refactoring correctness, but these metrics lack formal grounding in test comprehension. Our work introduces a test-specific complexity metric capturing structural signals—assertions, annotations, and test logic—ignored by existing syntactic metrics and embedding-based scores.

%% file: Sections/conclusion.tex
\section{Conclusion and Future Work}
\label{conclusion}

We introduced \textbf{CCTR}, a test-aware complexity metric designed to capture structural and semantic elements critical to unit test comprehension—such as assertion density, mocking constructs, and annotation semantics. Grounded in empirical studies on test readability, CCTR extends traditional metrics by aligning more closely with developer intuition. Our evaluation across 15,750 test suites shows that CCTR effectively differentiates between generation styles and produces scores reflective of structural effort. By bridging complexity analysis and test-specific readability, CCTR opens new possibilities for evaluating, refactoring, and improving unit tests. Future work will focus on human-centered validation, adaptive weighting, and integration into LLM-based test generation workflows. \\